\documentclass[11pt]{article}

\usepackage{amsmath}
\usepackage{amssymb}
\usepackage{graphicx}
\usepackage{lscape}
\usepackage{mathrsfs}
\usepackage{enumerate}
\usepackage{multirow}
\usepackage{rotating}
\usepackage{wrapfig}
\usepackage{url}
\usepackage{subfigure}
\usepackage{bm}
\usepackage{ulem}

\def\aa{A\&A}

\def\mnras{MNRAS}

\begin{document}

\vspace*{0.5cm}

\noindent {\Large THE ICRF-3: STATUS, PLANS, AND PROGRESS ON THE NEXT\\[2pt] GENERATION INTERNATIONAL CELESTIAL REFERENCE FRAME}
\vspace*{0.1cm}

\noindent\hspace*{1.5cm} Z. MALKIN$^{^1}$, C.S. JACOBS$^2$, F. ARIAS$^3$, D. BOBOLTZ$^4$, J. B\"OHM$^5$, S. BOLOTIN$^6$,\\
\noindent\hspace*{1.5cm} G. BOURDA$^{7,8}$, P. CHARLOT$^{7,8}$, A. DE~WITT$^9$, A. FEY$^{10}$, R. GAUME$^{10}$,\\
\noindent\hspace*{1.5cm} R. HEINKELMANN$^{11}$, S. LAMBERT$^{12}$, C. MA$^{13}$, A. NOTHNAGEL$^{14}$, M. SEITZ$^{15}$,\\
\noindent\hspace*{1.5cm} D. GORDON$^6$, E. SKURIKHINA$^{16}$, J. SOUCHAY$^{12}$, O. TITOV$^{17}$\\
\noindent\hspace*{1.5cm} $^1$ Pulkovo Observatory, Pulkovskoe Sh. 65, St.~Petersburg 196140, Russia\\
\noindent\hspace*{1.5cm} e-mail: malkin@gao.spb.ru\\
\noindent\hspace*{1.5cm} $^2$ Jet Propulsion Laboratory, CalTech/NASA, Pasadena, CA, USA\\
\noindent\hspace*{1.5cm} e-mail: Christopher.S.Jacobs@jpl.nasa.gov\\
\noindent\hspace*{1.5cm} $^3$ Bureau International des Poids et Mesures (BIPM), Paris, France\\
\noindent\hspace*{1.5cm} $^4$ Astronomical Sciences, National Science Foundation, Arlington, VA\\
\noindent\hspace*{1.5cm} $^5$ Technische Universit\"at Wien, Austria\\
\noindent\hspace*{1.5cm} $^6$ NVI, Inc./NASA Goddard Space Flight Center, Greenbelt, MD, USA\\
\noindent\hspace*{1.5cm} $^7$ Universit\'e de Bordeaux, LAB, UMR 5804, Floirac, France\\
\noindent\hspace*{1.5cm} $^8$ CNRS, LAB, UMR 5804, Floirac, France\\
\noindent\hspace*{1.5cm} $^9$ Hartebeesthoek Radio Astronomy Observatory, South Africa\\
\noindent\hspace*{1.5cm} $^{10}$ U.S. Naval Observatory, Washington D.C., USA\\
\noindent\hspace*{1.5cm} $^{11}$ Deutsches GeoForschungsZentrum Potsdam, Germany\\
\noindent\hspace*{1.5cm} $^{12}$ Observatoire de Paris, SYRTE, CNRS, UPMC, Paris, France\\
\noindent\hspace*{1.5cm} $^{13}$ NASA Goddard Space Flight Center, Greenbelt, MD, USA\\
\noindent\hspace*{1.5cm} $^{14}$ Institute of Geodesy and Geoinformation, University Bonn, Germany\\
\noindent\hspace*{1.5cm} $^{15}$ Deutsches Geod\"atisches Forschungsinstitut (DGFI), Munich, Germany\\
\noindent\hspace*{1.5cm} $^{16}$ Institute of Applied Astronomy, St.~Petersburg, Russia\\
\noindent\hspace*{1.5cm} $^{17}$ Geoscience Australia, Canberra, Australia\\

\footnotetext{Proceedings of the Journ\'ees 2014 ``Syst\`emes de r\'ef\'erence spatio-temporels'',
St. Petersburg, Russia, 22-24 Sep 2014, Eds. Z. Malkin, N. Capitaine, 2015, pp.~3--8.}

\vspace*{0.5cm}

\noindent {\large ABSTRACT.}
The goal of this presentation is to report the latest progress in creation of the next generation of VLBI-based International Celestial Reference
Frame, ICRF3.
Two main directions of ICRF3 development are improvement of the S/X-band frame and extension of the ICRF to higher frequencies.
Another important task of this work is the preparation for comparison of ICRF3 with the new generation optical frame GCRF expected by the end of the
decade as a result of the Gaia mission.

\vspace*{1cm}

\noindent {\large 1. INTRODUCTION}

\smallskip

In 1997, the International Celestial Reference Frame (ICRF) based on the positions of 608 extragalactic radio sources derived
from the VLBI observations at S/X bands has been adopted by the IAU as the fundamental celestial reference frame,
replacing the FK5 optical frame (Ma et al., 1998).
The first ICRF, hereafter referred to as ICRF1, was replaced in 2009 by ICRF2 also based on S/X observations (Ma et al., 2009), the current IAU
standard celestial reference frame.
The ICRF2 is very much improved with respect to ICRF1 in the sense of both number of sources included and position accuracy.
However, it still has serious problems discussed in Section 2.
To mitigate these problems, a new generation frame, the ICRF3, is currently under development making use of
both new VLBI observations and new developments in data analysis.
This work is coordinated by the IAU Division A Working Group Third Realization of International Celestial Reference Frame (Chair Christopher Jacobs).
We present here the current status of the ICRF3 as of September 2014 and prospects for the near future.

There are three primary tasks of the ICRF3 activity.
The first goal is a substantial improvement of ICRF2 in S/X band. The progress in this direction is described in Section 3.
The second task is to extend the ICRF to higher frequencies, such as Ka, K, and Q bands, which is crucial for many important practical
applications.
The third goal is to prepare the link of the new generation Gaia-based optical frame GCRF to ICRF3 by the end of the decade.
This problem is discussed in Section 5.

\vspace*{0.7cm}

\noindent {\large 2. CURRENT ICRF STATUS}

\smallskip

\begin{figure}[ht]
\centering
\vskip 2em
\includegraphics[width=0.52\textwidth,angle=-90,clip]{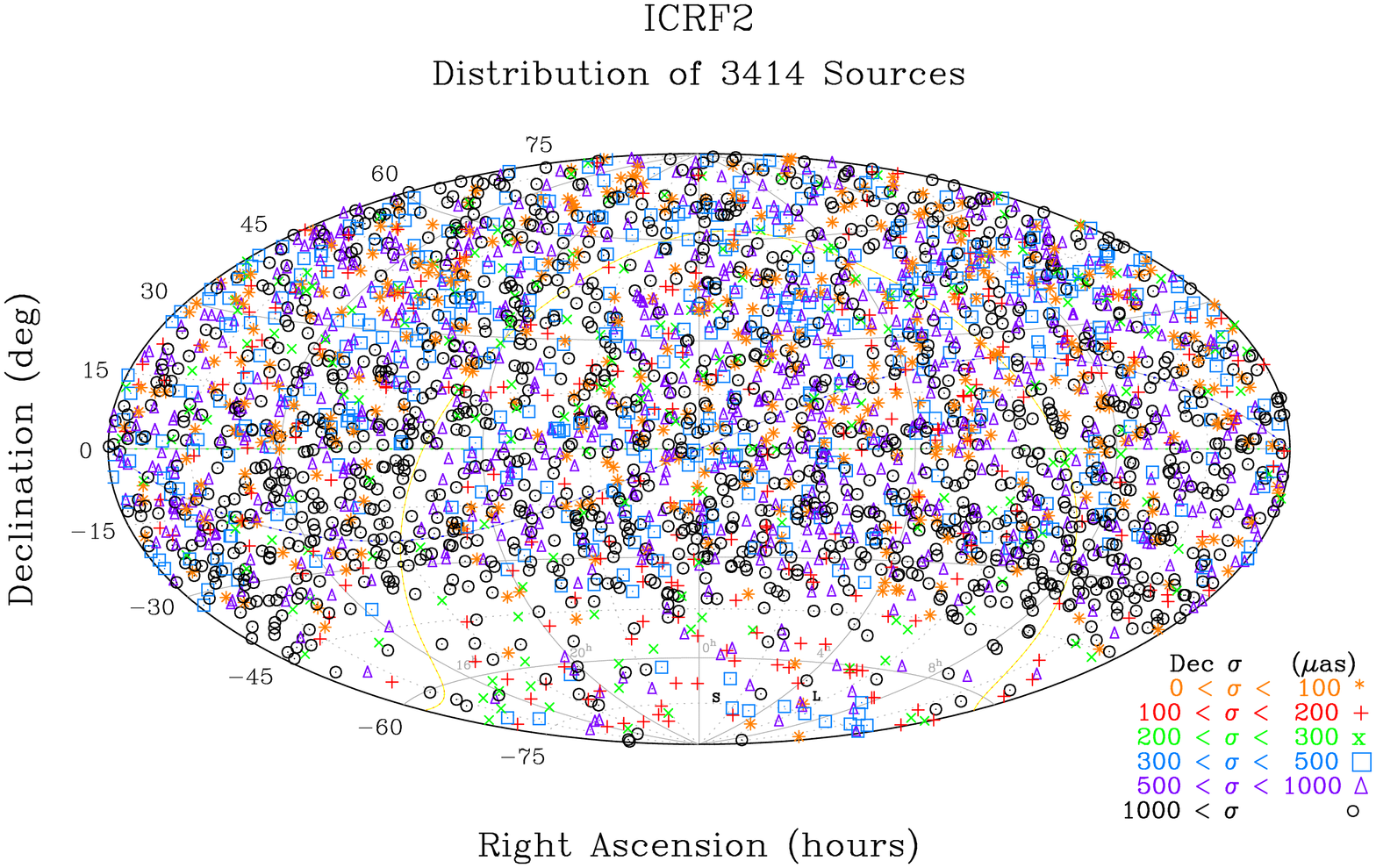}
\caption{ICRF2: the current IAU standard frame consists of 3414 sources (Ma et al., 2009). Note the
lower spatial density of sources south of $-30^\circ$. About 2/3 of the sources (2197) originating from the VCS survey
have 5 times lower precision than the well observed sources.}
\label{malkin_fig:ICRF2}
\end{figure}

\begin{figure}[ht]
\centering
\includegraphics[width=0.9\textwidth,clip]{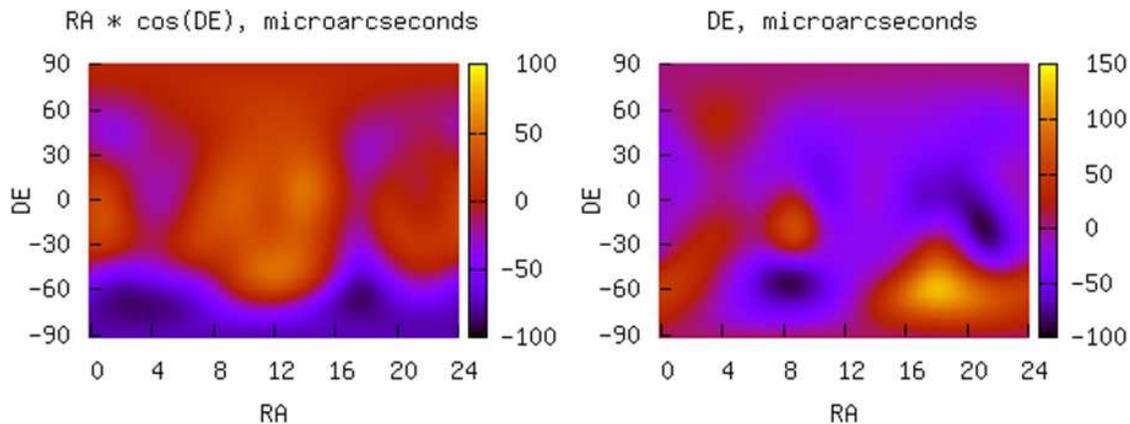}
\caption{ICRF2 minus ICRF1 smoothed  differences, $\mu$as.}
\label{malkin_fig:ICRF2-ICRF1}
\end{figure}

The ICRF2 catalog was computed using nearly 30 years of VLBI observations and provides accurate positions of 295 ``defining'' sources
and generally less accurate positions of 3119 other radio sources (Fig.~\ref{malkin_fig:ICRF2}).
The advantages of the ICRF2 with respect to the ICRF1 are manyfold:
\begin{itemize}
\itemsep=-0.7ex
\item increasing total number of sources from 608 (717 with two extensions) to 3414;
\item increasing number of the defining sources from 212 to 295 and improving their sky distribution;
\item more uniform distribution of the defining sources;
\item improving the source position uncertainty; decreasing the noise floor from 250~$\mu$as to 40~$\mu$as;
\item elimination of large ICRF1 systematic error at the level of $\approx$0.2 mas (Fig.~\ref{malkin_fig:ICRF2-ICRF1});
\item improving axes stability from $\approx$20~$\mu$as to $\approx$10~$\mu$as.
\end{itemize}

\begin{figure}[ht]
\centering
\includegraphics[width=0.6\textwidth,clip]{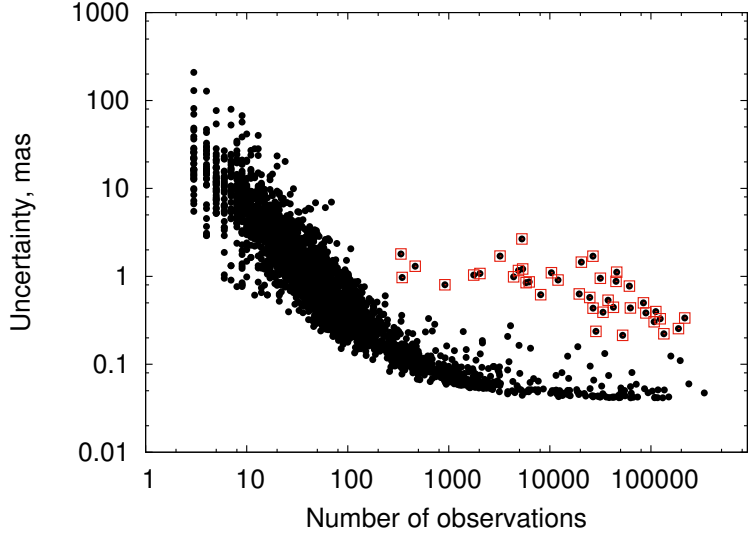}
\caption{Uncertainties of the ICRF2 source positions vs. number of observations. Note that the arc sources (highlighted) do not follow the general law.}
\label{malkin_fig:sigma_nobs}
\end{figure}

\begin{figure}[ht]
\includegraphics[width=0.7\textwidth,clip]{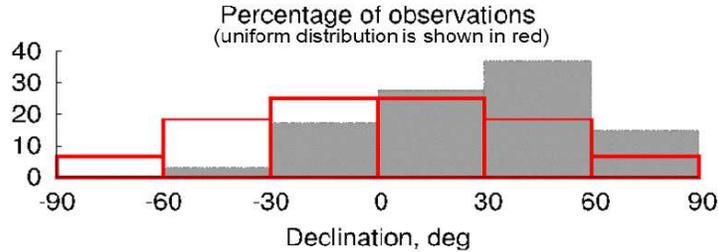}
\centering
\caption{Number of observations by declination bands. Note that the percentage of observations in the south polar cup region is not improved w.r.t. ICRF1.}
\label{malkin_fig:nobs_debands}
\end{figure}

\begin{figure}[ht]
\centering
\includegraphics[width=0.8\textwidth,clip]{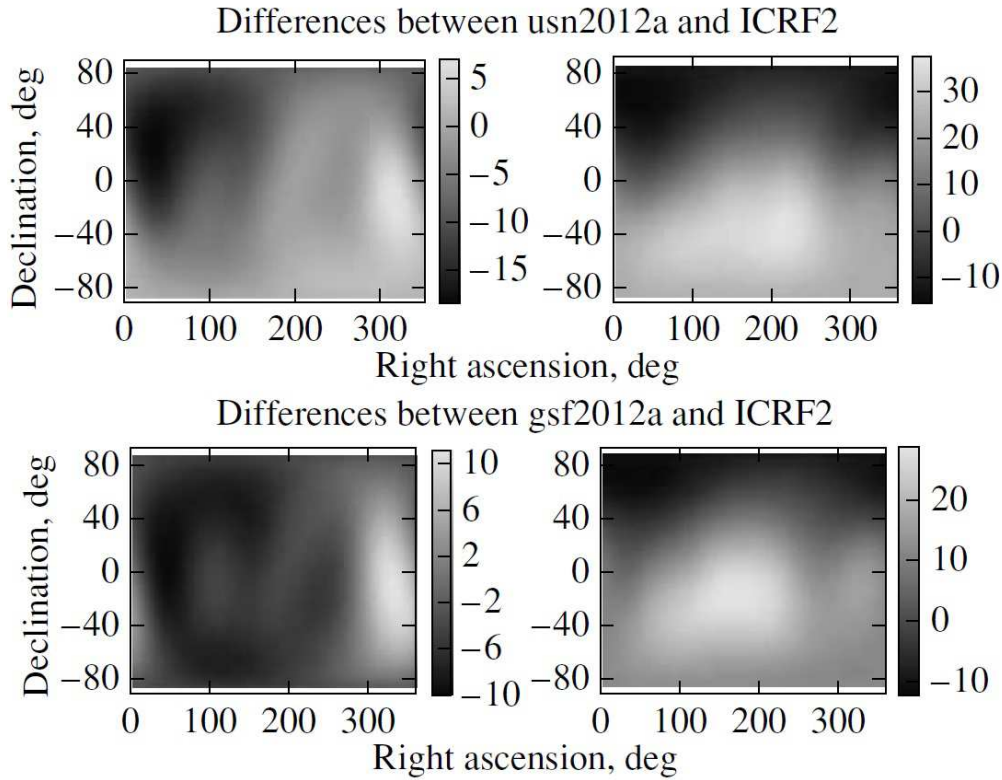}
\caption{Differences between recent VLBI catalogues and ICRF2, $\mu$as (Sokolova, Malkin, 2014).}
\label{malkin_fig:cat-ICRF2}
\end{figure}

\begin{figure}[ht]
\centering
\includegraphics[width=0.6\textwidth,clip]{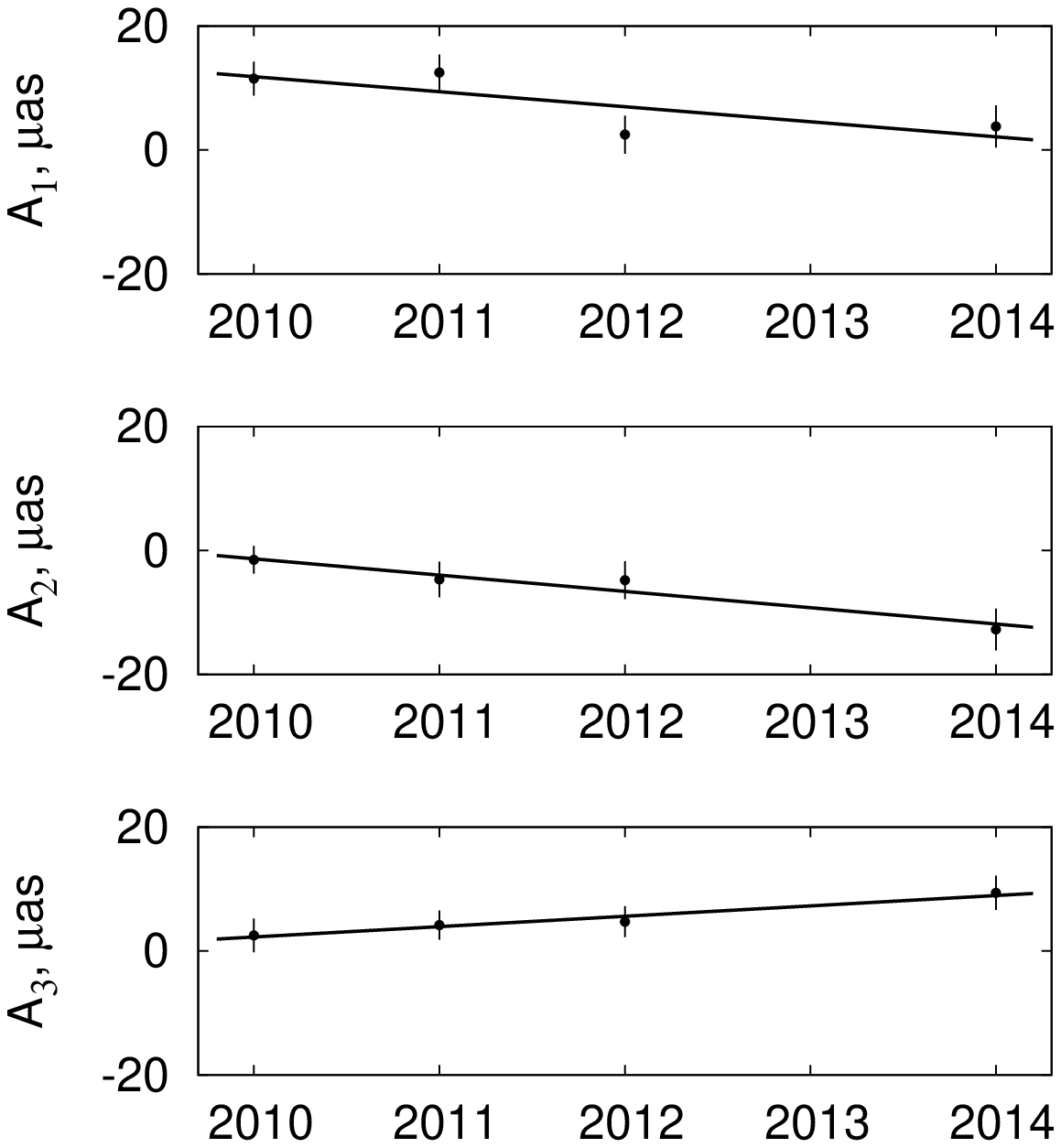}
\caption{Rotation of GSFC astrometric catalogues w.r.t. ICRF2 (Malkin, 2014).}
\label{malkin_fig:ICRF2rot}
\end{figure}

However, ICRF2 still has several serious deficiencies, the main of which are:
\begin{itemize}
\itemsep=-0.7ex
\item Very non-uniform distribution of the position accuracy. About 2/3 of the sources are from the VCS survey (Beasley et al., 2002) and have about
5 times worse median precision as compared with non-VCS ICRF2 sources.
Besides, 39 special handling unstable sources processed in {\it arc} mode have position uncertainties much large than other
sources having similar number of observations (Fig.~\ref{malkin_fig:sigma_nobs}).
\item Both, distribution of the ICRF2 sources and their position errors over the sky, are not uniform. Most of the sources are north of $-45^\circ$,
i.e. within VLBA sky coverage limits.
Due to the relatively small number of stations in the southern hemisphere (particularly a lack of large antennas),
position errors of the southern sources are generally substantially worse (Fig.~\ref{malkin_fig:ICRF2}).
In spite of putting into operation four new stations in Australia and New Zealand, the percentage of observations of southern sources, especially
in the southern polar cap region remains practically the same as for ICRF1 (Fig.~\ref{malkin_fig:nobs_debands}).
\item As follows from theoretical considerations (Liu et al., 2012) and analysis of the latest source position catalogs
(Malkin, 2014; Sokolova \& Malkin, 2014; Lambert, 2014), ICRF2 may have residual systematics at a level of $\approx$20~$\mu$as and rotations at a level
of a few $\mu$as per decade (Figs.~\ref{malkin_fig:cat-ICRF2}, \ref{malkin_fig:ICRF2rot}).
\item Official ICRF2 catalog is defined for S/X bands only, whereas many scientific and practical applications require the CRF realization of similar
quality for other frequencies.
\end{itemize}

The ICRF3 activity is aims at elimination of these ICRF2 problems taking advantage of gradually increasing total number of observations
(on average about 0.6~million observations per year during the last years), more active observations at southern stations, and new developments in
VLBI technology and data analysis.

\vspace*{0.7cm}

\noindent {\large 3. IMPROVING S/X ICRF}

\smallskip

The first problem to be solved for improving ICRF2 in S/X band is to achieve a more uniform distribution of the source position uncertainty.
Figure~\ref{malkin_fig:sigma_nobs} shows how it depends on the number of observations (dependence on the number of sessions is weaker).
Two main steps in this direction are now underway.

\begin{figure}[ht]
\centering
\includegraphics[width=\textwidth,clip]{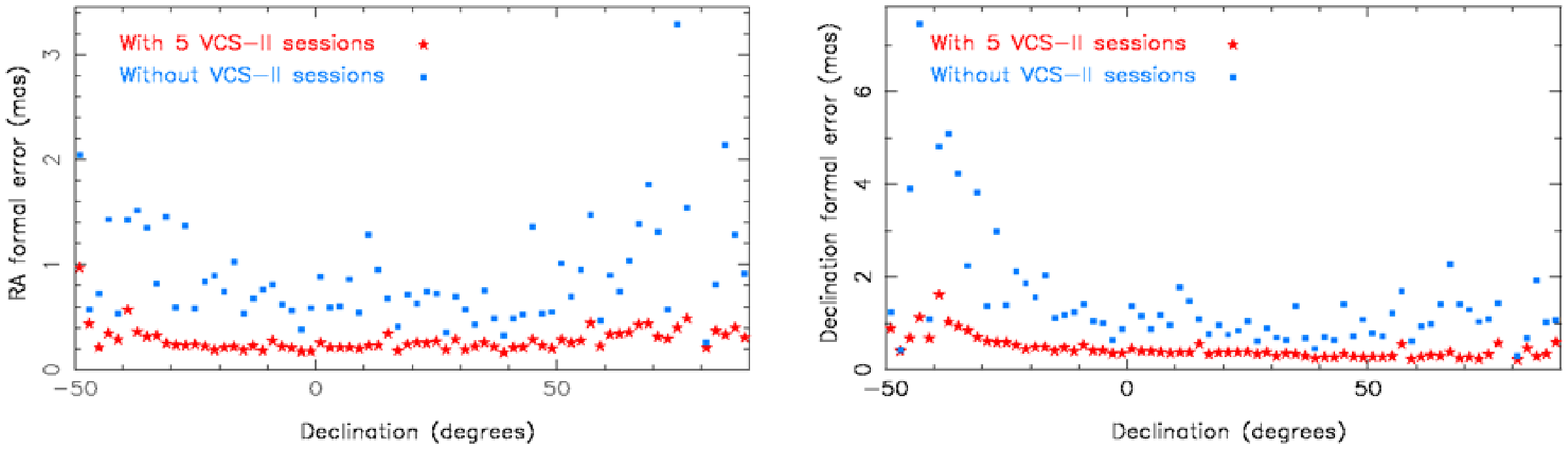}
\caption{VCS2: Average uncertainties in 2-degree bins for 1309 re-observed VSC sources.
 Note $\approx$3 times improvement in precision and much more uniform distribution of the position uncertainties
 over declination.}
\label{malkin_fig:VCS2}
\end{figure}

The VCS2 project was proposed and accepted by NRAO in 2014 (P.I. David Gordon).
Eight 24~h observing sessions are planned, and five of them have been observed, correlated, and analyzed at the GSFC VLBI group.
The first results of the analysis have shown manyfold improvement in the position uncertainty for re-observed VCS sources (Fig.~\ref{malkin_fig:VCS2}).

Improving the ICRF in the southern hemisphere in the sense of both the number of sources and their position accuracy is another primary task of the ICRF3.
A giant step in this direction was made with the inclusion of new VLBI antennas in Australia (Hobart, Katherine, Yarragadee),
New Zealand (Warkworth), and S. Africa (HartRao) in the IVS observing programs.
Because the new stations are equipped with relatively small antennas (12~m in Australia and New Zealand, and 15~m in S. Africa), larger antennas
such as Parkes 64~m, DSS45 34~m, Hobart 26~m, and HartRAO 26~m will need to be added in order to detect weaker sources (Titov et al., 2013).
Further improvement in the number of observations of southern sources can be achieved through inclusion of CRF sources in the regular IVS
EOP observing programs (Malkin et al., 2013).

Important factors limiting the source position precision and accuracy are source structure and the core-shift effect.
They are most significant in S/X band.
Both increasing of the number of many-baseline observations and developments in VLBI technology and analysis are needed to mitigate these effects.

~\vskip 0.2cm

\noindent {\large 4. EXTENDING ICRF TO HIGHER FREQUENCY BANDS}

\smallskip

As radio frequencies increase, sources tend to become more core dominated as the extended structure
in the jets tends to fade away with increasing frequency. Also the spatial offset of the radio emissions
from the AGN's central black hole due to opacity effects (core shift) is reduced with increasing observing
frequency. 

On the other hand, observations at K and Ka bands are more weather sensitive, which combined with the shorter wavelengths
leads to shorter coherence times. Furthermore, sources are often weaker and antenna pointing is more
difficult. The combined effect is lower sensitivity, but advances in recording technology are rapidly
compensating with higher data rates. Currently, the IVS, the VLBA and JPL's Deep Space Network are moving to 2~Gbps operations.

Currently, active CRF works are underway at K (22--24~GHz), Ka (32~GHz), and to a lesser extent Q (43~GHz) bands.

\begin{figure}[ht]
\centering
\includegraphics[width=0.52\textwidth,angle=-90,clip=true]{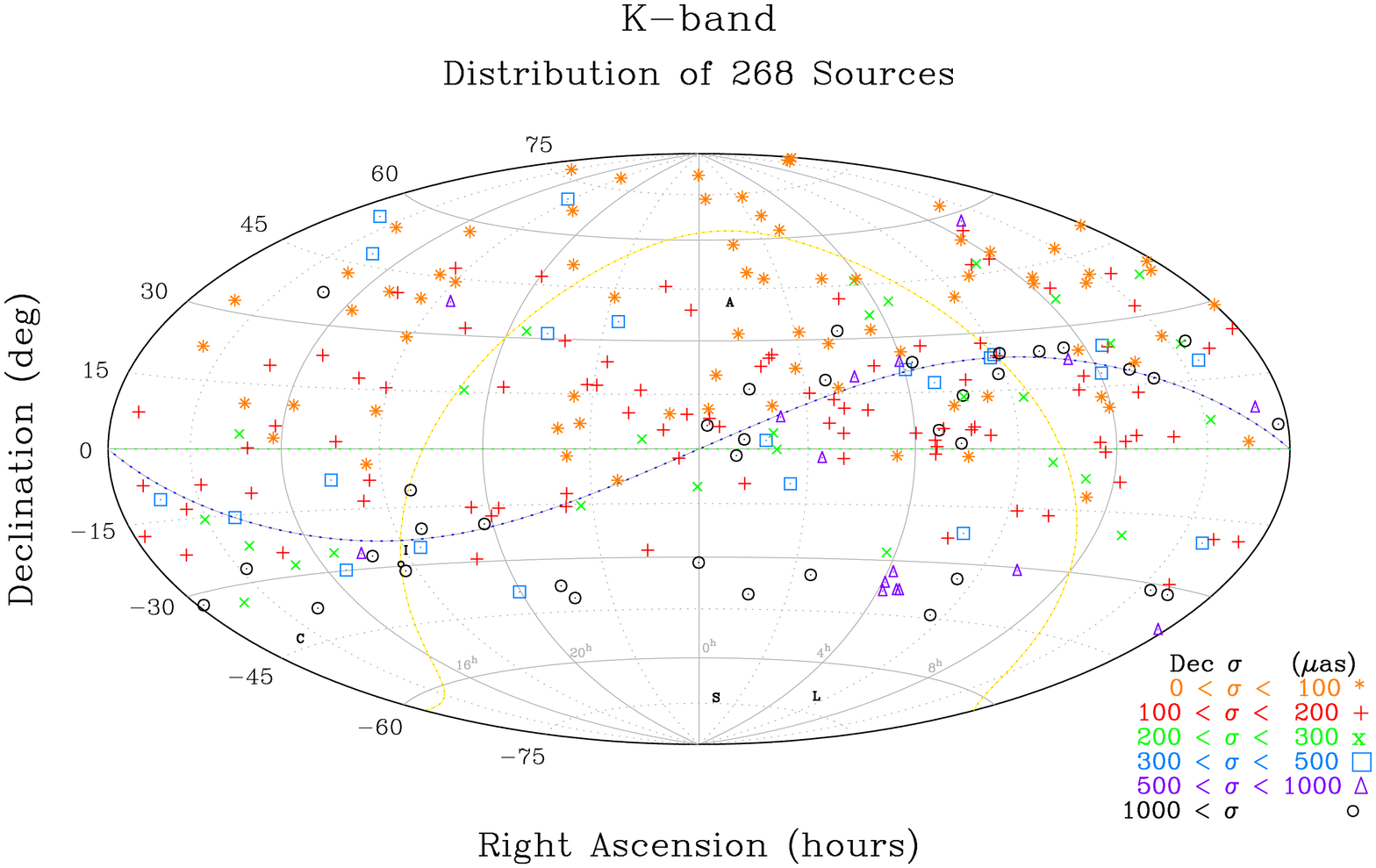}
\caption{K-band CRF: 268 sources, still weak in the south.}
\label{malkin_fig:k}
\end{figure}

Lanyi et al. (2010) and Charlot et al. (2010) did pioneering work to develop high precision celestial frames at 24 GHz.
Currently, the K-band CRF includes 275 sources (Fig.~\ref{malkin_fig:k}).
Most sources have a position precision better than 200 $\mu$as.
Further development is expected in the framework of activity of a new K-band full sky coverage collaboration (de Witt et al., 2014).
Accurate positions of more than 500 K-band sources are expected in the near future.

\begin{figure}[ht]
\centering
\includegraphics[width=0.52\textwidth,angle=-90,clip]{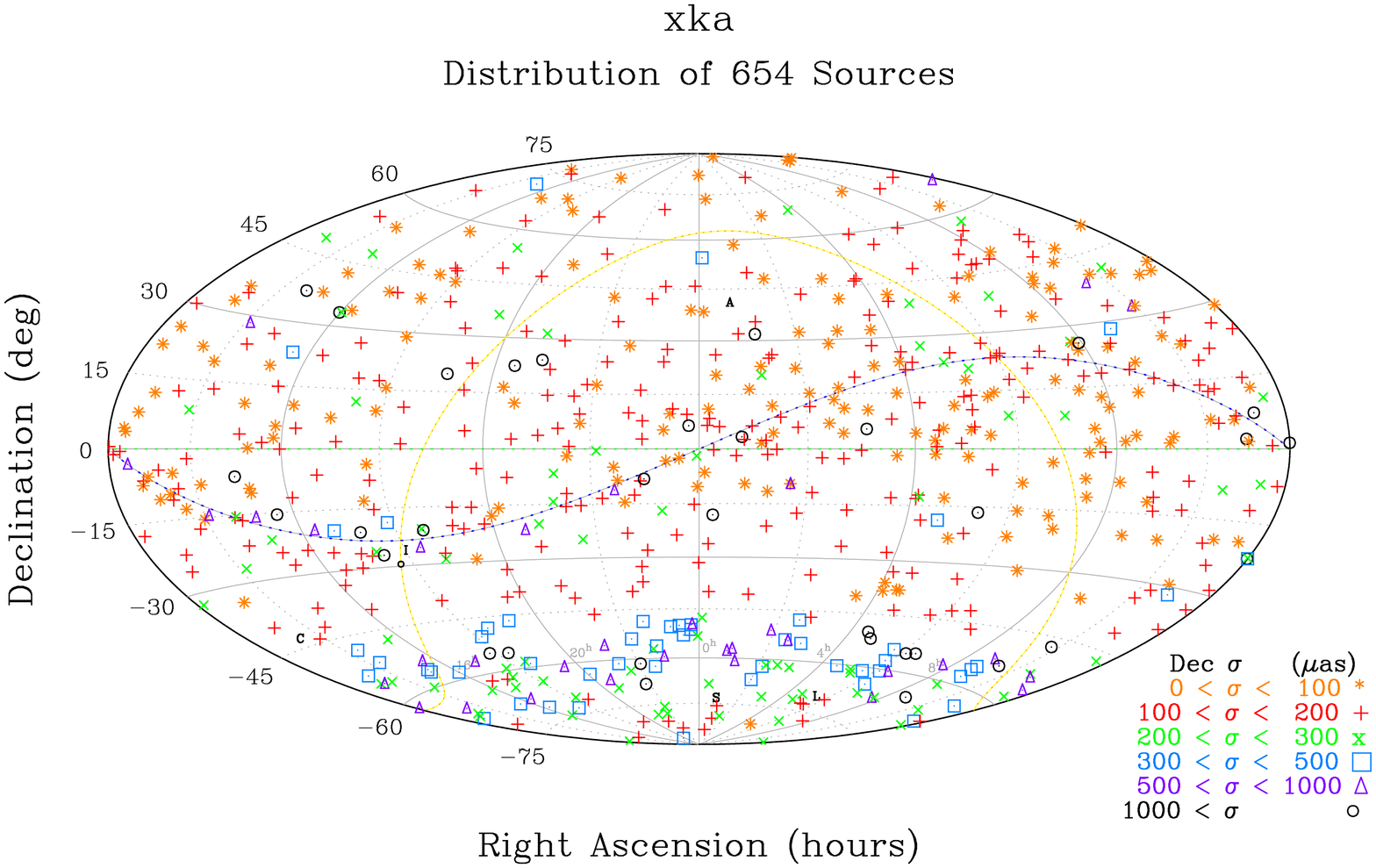}
\caption{X/Ka-band CRF: 654 sources, still weak in the south.}
\label{malkin_fig:ka}
\end{figure}

Since 2005, the two baselines of NASA's Deep Space Network have been making observations at X/Ka-band of about 500 sources
down to $-45^\circ$.
Recently they have been joined by ESA's DSA03 35~m antenna in Malarg\"ue, Argentina resulting in full sky coverage at Ka-band
(Horiuchi et al., 2013).
Now, the regularly observed Ka network consists of four stations: Goldstone (CA, USA), Tidbinbilla (A.C.T., Australia), Malarg\"ue (Mendoza, Argentina),
and Robledo de Chavela (Spain).
The current X/Ka-band CRF includes 644 sources (Fig.~\ref{malkin_fig:ka}), 700+ sources are expected in the near future.

It should be noted that, as with S/X, high frequency CRFs are still weak in the south.
\medskip

\vspace*{0.7cm}

\noindent {\large 5. OPTICAL--RADIO FRAMES LINK}

\smallskip

Launched in December 2013, ESA's Gaia mission is designed to make state-of-the-art astrometric
measurements (positions, proper motions and parallaxes) of a billion objects as well as photometric
and radial velocity measurements (Lindegren, 2008; Mignard, 2014).
Gaia's observations will include approximately 500,000 AGN of which $\approx$20,000 will be optically bright ($V < 18$ mag) thus enabling very
high expected precisions of 70--150~$\mu$as at $V=18^m$ and 25--50~$\mu$as at $V=16^m$.

Gaia Celestial Reference Frame (GCRF) will be created in two steps.
First an internally consistent solution will be computed from the data collected during the Gaia mission by the end of the decade.
Then this solution should be oriented in such a way to be consistent with the VLBI-based ICRF.

To provide the ICRF-GCRF link with the highest accuracy, dedicated efforts are underway in the framework of the ICRF3 activity.
First it is necessary to identify a sufficient number of optically and radio bright objects, whose positions can be reliably determined 
from both VLBI and Gaia observations with accuracy better than 100~$\mu$as.
Bourda et al. (2010) estimated that 300+ AGN are optically bright while also strong and compact in radio thus enabling both Gaia and VLBI to make
very precise position measurements.
This common set of sources should allow the GCRF and ICRF radio frames to be rotationally aligned to better than 10~$\mu$as precision.
After making the optical-radio alignment, position offsets between the two techniques can be studied to
characterize systematic errors. Having multiple radio frames (S/X, K, X/Ka) should be of great value in
characterizing frequency dependent effects e.g. core shift.

The work to extend the list of common Gaia-VLBI sources through the optical photometry of the current and prospective ICRF sources is underway
(Taris et al., 2013).

\vspace*{0.7cm}

\noindent {\large 6. CONCLUSIONS}

\smallskip

Our goals are to improve the precision, spatial and frequency coverage relative to the ICRF2 by 2018.
This date is driven by the desire to create radio frames that are ready for comparison with the Gaia optical frame.
Several specific actions are underway.
The VCS2 project is aimed at substantial improvement in S/X-band precision of about 2200 VCS sources.
Five sessions (of eight planned) are completed, and the first results are very encouraging.
S/X-band southern precision improvements are planned from observations with five new southern antennas, such as AuScope and HartRAO.
Both these factors: completion of the VCS2 and substantial increase of the number of astrometric VLBI observations
(currently about 9.8 million delays compared to 6.5 million delays used to derive ICRF2), especially in the south, makes it possible to publish
an intermediate ICRF version in 2015, which can be substantially improved with respect to ICRF2 and may be very useful for different
applications.

Large progress is also being achieved in developing the CRF at Ka and K bands.
New improvements are expected, in particular, from adding a new ESA station in Malarg\"ue, Argentina
thus providing three additional baselines to Australia, California and Spain.

On the analysis front, special attention will be given to combination techniques both of VLBI catalogs and of multiple data types
(Iddink et al., 2014, 2015; Seitz et al., 2014; Sokolova \& Malkin 2014).
Consistency of CRF, TRF, and EOP is another area of concern, see, e.g., Seitz et al. (2014).

The creation of a next generation VLBI-based ICRF and the Gaia-based new-generation optical GCRF are main projects
in fundamental astrometry for this decade.
Both frames are intended to provide ICRS realizations with systematic accuracy better than 10~$\mu$as.
It is anticipated that further comparison and merging of both the radio ICRF and the optical GCRF will allow construction of a new highly accurate
multiband and systematically uniform ICRF.

\bigskip\noindent\textit{Acknowledgements.}
The authors thank the International VLBI Service for Geodesy and Astrometry (IVS, Schuh \& Behrend, 2012) and its members for decades
of dedication to the collection of the data used in this research.
This work is done in part under NASA contract.
Sponsorship by U.S. Government, as well as other respective institutes and funding agencies is acknowledged.

\vspace*{0.7cm}

\noindent {\large 7. REFERENCES}

{

\leftskip=5mm
\parindent=-5mm

\smallskip

Beasley, A.J., et al., 2002, ``VLBA Calibrator Survey'', ApJS, 141,  pp.~13--21.

de Witt, A., et al., 2014, ``Extending the K-band celestial frame emphasizing Southern hemisphere'', In: Proc. Journ\'ees 2013 ``Syst\`emes de R\'ef\'erence Spatio-Temporels'', N.~Capitaine (ed.), Observatoire de Paris, pp.~61--64.

Bourda, G., et al., 2010, ``VLBI observations of optically-bright extragalactic radio sources for the alignment of the radio frame with the future Gaia frame'', \aa, 520, A113.

Charlot, P., et al., 2010, ``The Celestial Reference Frame at 24 and 43 GHz. II. Imaging'', AJ, 139, pp.~1713--1770 .

Horiuchi, S., et al., 2013, ``The X/Ka Celestial Reference Frame: Results from combined NASA-ESA baselines'', Asia-Pacific Radio Astronomy Conference 2013, ADS: 2013apra.confE...1H.

Iddink, A., et al., 2014, ``Rigorous VLBI intra-technique combination strategy for upcoming CRF realizations'', , In: Proc. Journ\'ees 2013 ``Syst\`emes de R\'ef\'erence Spatio-Temporels'', N.~Capitaine (ed.), Observatoire de Paris, pp.~81--83.

Iddink, A., et al., 2015, ``First results of S/X and X/Ka-band catalogue combinations with full covariance information'', this volume, pp.~20--23.

Lambert, S., 2014, ``Comparison of VLBI radio source catalogs'', A\&A, 570, A108.

Lanyi, G., et al., 2010, ``The Celestial Reference Frame at 24 and 43 GHz. I. Astrometry'', AJ, 139, pp.~1695--1712.

Lindegren, L., et al., 2008, ``The Gaia Mission: Science, Organization and Present Status'', IAU Symp. 248, pp.~217--223.

Liu, J.-C., et al. 2012, ``Systematic effect of the Galactic aberration on the ICRS realization and the Earth orientation parameters'', A\&A, 548, A50.

Ma, et al., 1998, ``The International Celestial Reference Frame as Realized by Very Long Baseline Interferometry'', AJ, 116, pp.~516--546.

Ma, et al., 2009, ``The second realization of the International Celestial Reference Frame by Very Long Baseline Interferometry'', IERS Technical Note 35, Fey,~A.L., Gordon,~D., Jacobs,~C.S. (eds.). 

Malkin, Z., et al., 2013, ``Searching for an Optimal Strategy to Intensify Observations of the Southern ICRF sources in the framework of the
regular IVS observing programs'', In: Proc. 21st Meeting of the EVGA, N.~Zubko, M.~Poutanen (eds.), Rep. Finn. Geod. Inst., 2013:1, pp.~199--203.

Malkin, Z., 2014, ``On the implications of the Galactic aberration in proper motions for the Celestial Reference Frame'', \mnras, 445, pp.~845--849.

Mignard, F., 2014, ``Gaia status and early mission'', In: Proc. Journ\'ees 2013 ``Syst\`emes de R\'ef\'erence Spatio-Temporels'', N.~Capitaine (ed.), Observatoire de Paris, pp.~57--60.

Schuh, H., Behrend, D., 2012, ``VLBI: A fascinating technique for geodesy and astrometry'', J. Geodyn., 61, pp.~68--80.

Seitz, M., et al., 2014, ``Consistent adjustment of combined terrestrial and celestial reference frames'', In: Rizos C., Willis P. (eds.), IAG Symposia, 139, pp.~215--221. 

Sokolova, Y., Malkin, Z., 2014, ``Pulkovo combined catalogue of radio source positions PUL 2013'', Astron. Lett., 40, pp.~268--277.

Taris, F., et al., 2013, ``Optical monitoring of extragalactic sources for linking the ICRF and the future Gaia celestial reference frame. I. Variability of ICRF sources'', A\&A, 552, A98.

Titov, O., et al., 2013, ``International collaboration for improvement of the Celestial Reference Frame in the southern hemisphere'', IAG Symposium, Potsdam.
\\ \url{http://www.iag2013.org/IAG_2013/Publication_files/abstracts_iag_2013_2808.pdf}

}

\end{document}